\documentclass[aps,prl,reprint,superscriptaddress]{revtex4-2}
\usepackage{graphicx}
\usepackage{tabularx}
\usepackage{dcolumn}
\usepackage{longtable}
\usepackage{tensor}
\usepackage{color}
\usepackage{placeins}
\usepackage{bm}
\usepackage{amsmath}
\usepackage{amsfonts}
\usepackage{amssymb}
\usepackage[urlcolor=white]{hyperref}%
\providecommand{\U}[1]{\protect\rule{.1in}{.1in}}
\graphicspath{{./Figures/}}
\graphicspath{{./Figures/}}
\begin{document}
\title{Non-Perturbative Vibrational Excitation by Arbitrary Electric Fields from Classical Phase-Space Dynamics
}

\author{Sameernandan Upadhyayula}
\author{Ji\v{r}\'i J. L. Van\'i\v{c}ek}
\email{jiri.vanicek@epfl.ch \\}

\affiliation{Laboratory of Theoretical Physical Chemistry, Institut des Sciences et Ing\'enierie Chimiques, Ecole Polytechnique F\'ed\'erale de Lausanne (EPFL), CH-1015, Lausanne, Switzerland}

\date{\today}

\begin{abstract}
We present a non-perturbative framework for the quantum treatment of vibrational excitation in molecular systems driven by electric fields with arbitrary time dependence. For harmonic potentials with linear dipole coupling, the field-induced dynamics is exactly described by a displaced coherent state, whose evolution is completely determined by a complex phase-space coordinate. Closed-form expressions for this coordinate reveal a clear distinction between resonant and non-resonant regimes. The resulting vibrational populations follow a Poisson distribution determined by a dimensionless time-dependent phase-space displacement, for both resonant and non-resonant driving.  Resonant driving results in an enhanced phase-space displacement, whereas finite detuning between the driving and vibrational frequencies leads to an exponential suppression of the phase-space displacement and vibrational excitation. While the approach reduces to the standard time-dependent perturbation theory in the weak-field limit, it remains valid in the strong-field regime, where perturbative treatments fail. Our results provide a transparent phase-space picture of vibrational control and enable non-perturbative state preparation in ultrafast spectroscopy.
\end{abstract}

\maketitle
\begin{figure}[htbp]
\centering
\includegraphics[width=1.0\linewidth,  trim=0 60 0 60, clip]{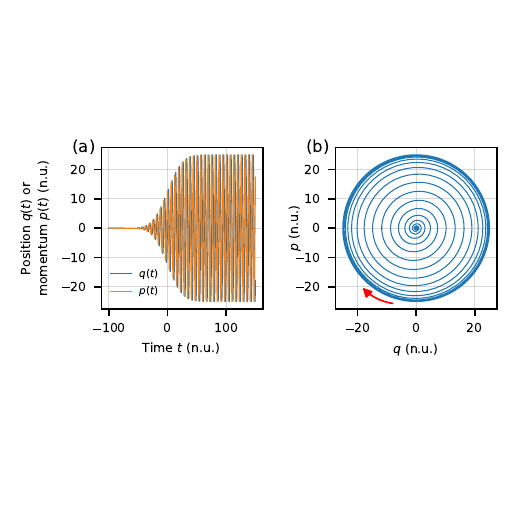}
\vspace{-1.0cm}
\caption{(a) Time evolution of position and momentum under a resonant electric field. (b) The corresponding phase-space trajectory showing a clockwise evolution (red arrow indicates the direction of motion). We perform the calculations in natural units (n.u.) such that $\hbar=\mu_0'=\omega_0=1$, $\mu_0=0$. The parameters of the electric field are    $E_0=\omega=1 \ \text{n.u.}$, $\varphi=0$, and $\sigma=20 \ \text{n.u.}$. The initial point is $(q_0,p_0)=(0,0)$. The asymptotic amplitude of $q$ and $p$ in the long-time limit obtained from the numerical evaluation is $25.066$ n.u., which agrees with Eq.~(\ref{13}). }
\label{fig:1}
\end{figure}
\begin{figure}[htbp]
\centering
\includegraphics[width=1.0\linewidth,  trim=0 66 0 66, clip]{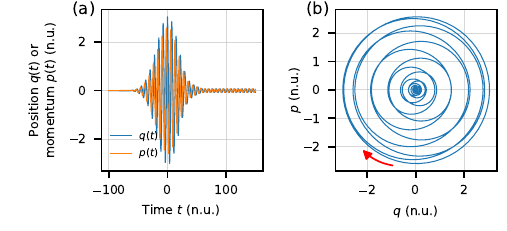}
\vspace{-1.0cm}
\caption{(a) Time evolution of position and momentum under a non-resonant electric field. (b) The corresponding phase-space trajectory, showing a clockwise evolution (red arrow indicates direction) but with no clear order in the structure of phase space. The parameters are the same as in Fig.~\ref{1} except that $\omega=0.83 \ \text{n.u.} \ (<\omega_0)$. In this case, $|\theta_-|\sigma=3.4$ is not very large, and therefore the trajectories show residual oscillations. The amplitude of $q$ and $p$ in the long-time limit obtained from the numerical evaluation is $0.0774$ n.u., which agrees with Eq.~(\ref{14}).}
\label{fig:2}
\end{figure}
\begin{figure}[htbp]
\centering
\includegraphics[width=1.0\linewidth,  trim=0 45 0 45, clip]{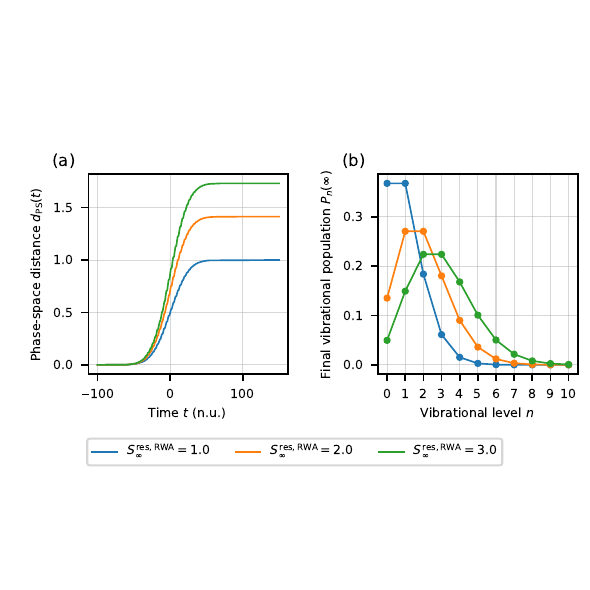}
\vspace{-1.0cm}
\caption{(a) Time evolution of the phase-space distance $d_{\text{PS}}(t)$ for different values of $S_{\infty}^{\text{res,RWA}}$ under resonant driving. (b) Corresponding final vibrational population distribution over number states. All parameters are identical to Fig.~\ref{1} except for the amplitude $E_0$, which is varied to control $S_{\infty}^{\text{res,RWA}}$.}
\label{fig:3}
\end{figure}
\begin{figure}[htbp]
\centering
\includegraphics[width=1.0\linewidth,  trim=0 52 0 52, clip]{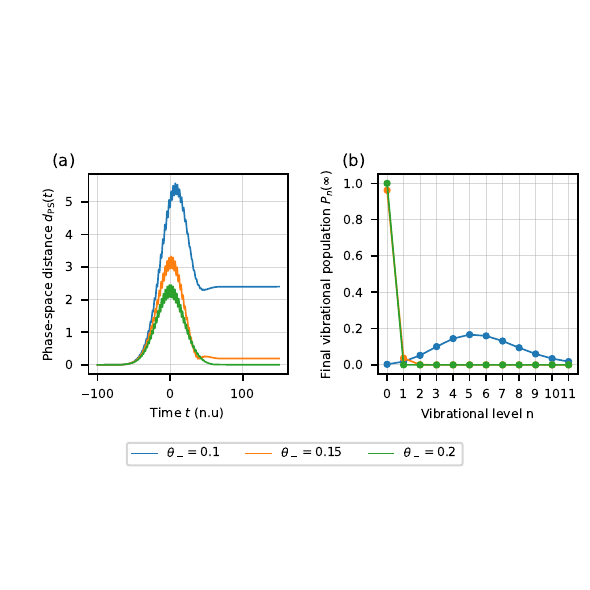}
\caption{(a) Time evolution of the phase-space distance $d_{\text{PS}}(t)$ in the non-resonant case for different values of the detuning parameter $\theta_-=|\omega-\omega_0|$. (b)  Corresponding final vibrational population distribution over number states. All parameters are identical to Fig.~\ref{2}, except for the frequency of the pulse, which is varied in order to vary detuning.    }
\label{fig:4}
\end{figure}
\begin{figure}[htbp]
\centering
\includegraphics[width=1.0\linewidth,  trim=0 55 0 55, clip]{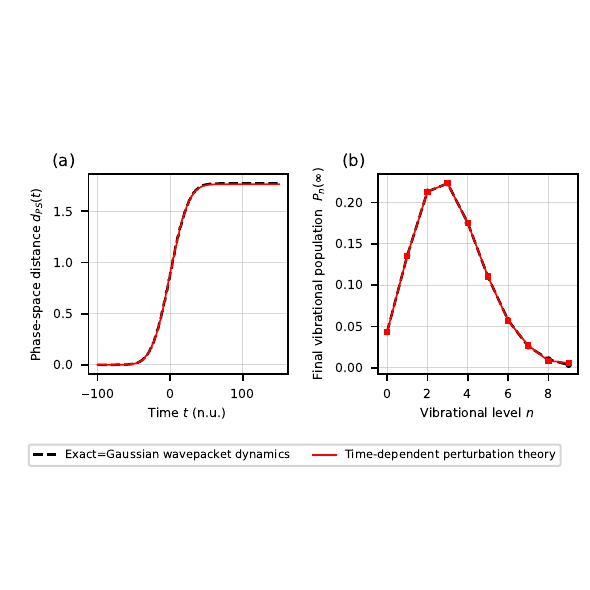}
\caption{Comparison between time-dependent perturbation theory and Gaussian wavepacket dynamics in the weak-field limit under resonant driving. (a) Time evolution of the phase-space distance. (b) Final vibrational populations. The parameters used for the calculations are the same as in Fig.~\ref{fig:1}, except  $E_0=0.1$ n.u. } 
\label{fig:5}
\end{figure}
\begin{figure}[htbp]
\centering
\includegraphics[width=1.0\linewidth,  trim=0 55 0 55, clip]{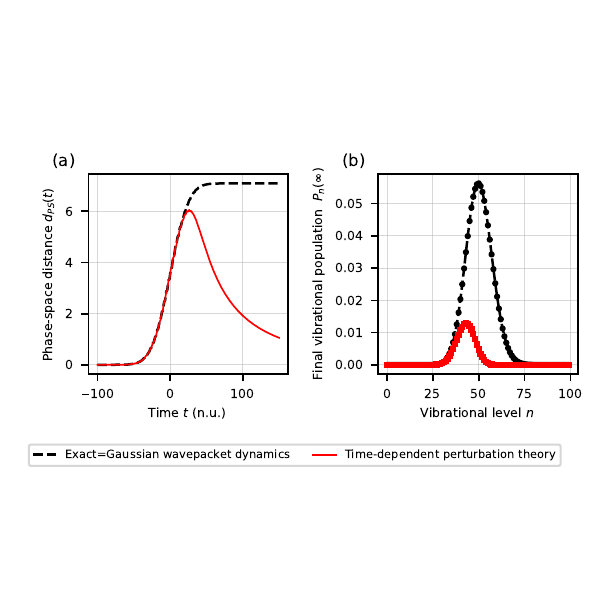}
\caption{Comparison between time-dependent perturbation theory and Gaussian wavepacket dynamics in the strong-field limit under resonant driving. (a) Time evolution of the phase-space distance. (b) Final vibrational populations. The parameters used for the calculations are the same as in Fig.~\ref{1}, except $E_0=0.4$ n.u.  } 
\label{fig:6}
\end{figure}

\clearpage
Vibrational excitation of molecular systems by infrared (IR) fields provides a direct route to controlling nuclear motion and energy flow \cite{Heberle2010,Zanni2011,Raschke2015}. IR radiation couples selectively to dipole-active normal modes, and the efficiency of excitation is governed not only by the carrier frequency of the field but also by its temporal envelope and phase, which together determine the spectral bandwidth and its overlap with molecular vibrational spectrum. In the time domain, this interaction induces coherent nuclear motion, while from a spectroscopic perspective it manifests as population transfer among vibrational states. In its simplest form, this interaction reduces to a driven harmonic oscillator, where excitation efficiency is governed by the spectral overlap between the field and the vibrational motion. Recent experiments exploit this idea by combining IR vibrational pre-excitation \cite{Bredenbeck2015,Khalil20152,Tokmakoff2016} with subsequent UV/VIS electronic excitation, notably in the Vibrationally Promoted Electronic Resonance (VIPER) scheme \cite{Bredenbeck2014,Bredenbeck2022}. In this approach, a selective IR pulse prepares a non-equilibrium vibrational state that enables mode-specific control of photochemical pathways \cite{Pascual2004,Evangelista2026}. Theoretical descriptions of vibrational excitation typically invoke the sudden approximation, in which the vibrationally excited state is prepared instantaneously, without considering the role of the explicit form of the IR field on the excitation process \cite{Burghardt2017,Burghardt2023}. Time-dependent perturbation theory (TDPT) provides an alternative but is accurate only when the net effect of the driving field remains small. This limitation becomes critical for strong fields, where higher-order effects govern the dynamics and perturbative treatments fail to capture the resulting vibrational populations. 
In this paper, we describe a non-perturbative phase-space framework for vibrational excitation, which provides an exact description for harmonic systems with linear dipole coupling and remains valid beyond the perturbative regime. \\
Let us consider a molecular system interacting with an external time-dependent electric field, described by a time-dependent Hamiltonian
\begin{align}
\hat H_{\text{tot}}(t) = \hat H_\text{vib}+ \hat V_{\text{int}}(t) \label{1},
\end{align}
where $\hat H_\text{vib}= T(\hat p)+ V_\text{vib}(\hat q)$ is the vibrational Hamiltonian expressed in mass-scaled normal mode coordinates, $T(\hat p)=\hat p^2/2$ is the vibrational kinetic energy and $\hat V_{\text{int}}(t) =-\hat{\vec{\mu}}\cdot \vec{E}(t)=-\hat\mu E(t)$ is the interaction potential energy within the electric-dipole approximation \cite{Ratner2002} 
with $\vec E(t)=\vec \epsilon E(t)$ denoting the external electric field with a unit polarization vector $\vec\epsilon$ and $\hat\mu=\hat{\vec{\mu}}\cdot\vec\epsilon$ the projection of the molecular electric dipole along $\vec\epsilon$.  To account for vibrational excitation, the molecular electric dipole operator is approximated by an affine function 
\begin{align}
\mu(q)&=\mu_0+{\mu_0'}^T\cdot q,\label{2} 
\end{align}
where $\mu_0'$ is the “gradient” of the dipole moment vector $\mu(q)$ with respect to the nuclear coordinates at the equilibrium geometry. Because IR-driven vibrational excitation is inherently mode-selective, we restrict the analysis to a single normal mode, which is equivalent to a diatomic molecule with a vibrational frequency $\omega_0$. Note, however, that the procedure described below can be easily generalized to polyatomic molecules. \\
Assuming $\hat H_\text{vib}$ to be harmonic and the classical dipole operator to be affine [as in Eq.~(\ref{2})], the total Hamiltonian $\hat H_{\text{tot}}(t)$ describes a harmonic oscillator driven by a time-dependent external field. The corresponding time-dependent Schr\"odinger equation 
\begin{align}
i\hbar|\dot\psi(t)\rangle=\hat H_{\text{tot}}(t)|\psi(t)\rangle,\label{3}
\end{align}
admits a displaced coherent-state solution \cite{Glauber1963,Sudarshan1965,Bargmann1971,Um1989,Kim1996}
\begin{align}
|\psi(t)\rangle=\exp\left[i\phi(t)/\hbar\right]\hat D(z_t)|\psi(-\infty)\rangle
\label{4}
\end{align}
where $\hat D(z_t)=\exp(z_t\hat a^\dag-z_t^\ast\hat a)$ is the displacement operator which operates on the initial wavepacket, while $\hat a^\dag$ and $\hat a$ are the harmonic-oscillator creation and annihilation operators, respectively. The time evolution of the coherent state is encoded in the complex phase-space coordinate  \cite{Quack1989,Tannor2007}
\begin{align}
z_t=\frac{1}{\sqrt{2\hbar\omega_0}}(\omega_0 q_t+ip_t),\label{5}
\end{align} 
determined by the corresponding classical trajectory with position $q_t$ and momentum $p_t$. The dynamical phase $\phi(t)/\hbar$ is given by the classical action $\phi(t):=\int_{-\infty}^t \left[p(t)\dot q(t)-H_{\text{tot}}(p_t,q_t)\right] dt$. \\
The displaced coherent-state solution in Eq.~(\ref{4}) is a particular realization of the more general Gaussian wavepacket, described in position representation by the wavefunction
\begin{equation}
\psi(q,t)=\exp\left[\frac{i}{\hbar}\left(\frac{1}{2}A_t x^2+p_t x+\gamma_t\right)\right],
\label{6}
\end{equation}
where $x=q-q_t$ is the position relative to the instantaneous wavepacket center and $q_t$, $p_t$, $A_t$, and $\gamma_t$ are  time-dependent parameters \cite{Heller1975,Heller1976,Vanicek2023}. The complex number $A_t$ (which becomes a symmetric matrix in polyatomic molecules) determines the width of the wavepacket through its imaginary part and the associated spatial chirp through its real part, while the complex number $\gamma_t$ accounts for the overall phase and normalization. For a harmonic vibrational mode, the coherent state corresponds to the special case of the Gaussian wavepacket with a time-independent width equal to that of the vibrational ground state, implying $A_t=A_{-\infty}=i\omega_0$, while the wavepacket center $(q_t,p_t)$ determines the real and imaginary parts of the complex phase-space coordinate $z_t$. Thus, the coherent-state solution employed here is equivalent to a  “frozen” Gaussian wavepacket \cite{Heller1981}, whose parameters evolve according to the equations
\begin{equation}
\begin{aligned}
\dot q_t &= p_t,\\
\dot p_t &= -V_{\mathrm{vib}}'(q_t)+\mu_0'E(t),\\
A_t &= \text{constant}=i\omega_0,\\
\dot\gamma_t &= T(p_t)-V_{\mathrm{vib}}(q_t)
-\hbar\omega_0/2
+\mu_0 E(t).
\end{aligned}
\label{7}
\end{equation}
Taking the modulus squared of the overlap between the propagated coherent state in Eq.~(\ref{4}) and the $n$th vibrational eigenstate yields the exact, time-dependent population of the $n$th vibrational level; these populations follow the time-dependent Poisson distribution
\begin{align}
P_n(t)=\exp(-S_t){S_t^{n}}/{n!},\label{8}
\end{align}
determined by a single dimensionless parameter $S_t=|z_t|^2$, equal to the square of the dimensionless phase-space distance, which thus controls both the mean and variance of the vibrational spectrum.
\\
The complex phase-space coordinate for this simplified model can be evaluated as
\begin{align}
z_t=\frac{i\exp(-i\omega_0t)}{\sqrt{2\hbar\omega_0}}\int_{-\infty}^t\exp(i\omega_0s)\left[\mu_0'E(s)\right] \ ds. \label{9}
\end{align}
To obtain a closed-form solution for Eq.~(\ref{9}), we consider an electric field of the form 
\begin{align}
E(t)=E_0\frac{d}{dt}\left[g(t)\sin(\omega t-\varphi)\right] \ , \ g(t)=\exp\left(-{t^2}/{2\sigma^2}\right),\label{10}
\end{align}
where $E_0$ is the field amplitude, and $\omega,\ \sigma$ define its carrier frequency and temporal width. The structure of the electric field defined in Eq.~(\ref{10}) describes a realistic short pulse with a chirp and approximately a Gaussian envelope and ensures that its time integral from -$\infty$ to $\infty$ is 0. This condition guarantees that the corresponding vector potential is the same before and after the pulse and that the field does not impart any momentum to the system \cite{Jackson1999}. Substitution into Eq.~(\ref{9}) yields a closed-form expression for the phase-space coordinate, 
\begin{multline}
z_{t} =
\frac{i(\mu_0'E_{0})}
{\sqrt{2\hbar\omega_0}}
\Bigg\{
g(t)\sin(\omega t-\varphi) \\
-{\frac{\omega_0}{2} \exp(-i\omega_0 t)}
\left[
\exp(-i\varphi)J(\theta_+,t)
-
\exp(i\varphi)J(\theta_-,t)
\right]
\Bigg\}, \label{11}
\end{multline}
where the functions $J(\theta_\pm,t)$ are given by
\begin{align}
J(\theta_{\pm},t)&=\sqrt{\frac{\pi}{2}}\sigma\exp\left(-\frac{\theta_{\pm}^2\sigma^2}{2}\right)\left[1+\operatorname{erf}\left(\frac{t}{\sqrt{2}\sigma}+\frac{i\theta_\pm\sigma}{\sqrt{2}}\right)\right]. \label{12}
\end{align}
Equation (\ref{11}) separates the dynamics into components oscillating with angular frequencies $\theta_\pm=\omega\pm\omega_0$, corresponding to co-rotating ({-}) and counter-rotating ({+}) contributions of the driving field. Their relative weight is controlled by the Gaussian factors $\exp\left(-\frac{\theta_\pm^2\sigma^2}{2}\right)$, which directly encode the spectral overlap between the pulse and vibrational frequency. 
\\
At resonance $(\omega=\omega_0)$, the counter-rotating contribution $J(\theta_+,t)$ is exponentially suppressed for $\omega_0\sigma\gg1$, due to its rapid oscillations and negligible spectral overlap. This justifies the rotating-wave approximation (RWA) \cite{Bloch1940,CohenTannoudji1992,Vanicek2014,Grossmann2020,Grossmann2026}, in which only the co-rotating term $J(\theta_-,t)$ contributes to the dynamics. In the long-time limit, after the pulse has vanished, $z_t$ simplifies to
\begin{align}
z_{\infty}^{\text{res,RWA}}=\frac{i\sigma\left(\mu_0'E_{0}\right)}{2}\sqrt{\frac{\pi\omega_0}{\hbar}}\exp[-i(\omega_0 t-\varphi)].\label{13}
\end{align}
Thus, after the pulse has ended, the system undergoes a free evolution with a constant amplitude. The associated phase-space picture of the excitation corresponds to an outward spiral converging to a limit cycle (Fig.~\ref{fig:1}).
\\
In contrast, for non-resonant driving, the long-time trajectory is given by
\begin{equation}
\begin{aligned}
z_{\infty}^{\mathrm{non-res}}
={}&-\frac{i\sigma\mu_0'E_0}{2}
\sqrt{\frac{\pi\omega_0}{\hbar}}
\Bigg\{
\exp\left[-\frac{\theta_+^2\sigma^2}{2}
-i(\omega_0t+\varphi)\right] \\
&\qquad
-\exp\left[-\frac{\theta_-^2\sigma^2}{2}
-i(\omega_0t-\varphi)\right]
\Bigg\}.\label{14}
\end{aligned}
\end{equation}
For typical pulses satisfying $\theta_+\sigma\gg 1$, the counter-rotating contribution is exponentially suppressed, and one can still use the rotating-wave approximation even when $|\theta_-|\sigma\gg1$. The amplitude is governed by the detuning parameter $\theta_-=\omega-\omega_0$, resulting in an exponentially small asymptotic phase-space displacement. In contrast to the resonant case, the non-resonant phase-space dynamics exhibits much more residual interference between the co-rotating and counter-rotating contributions, which leads to a less regular phase-space trajectory (Fig.~\ref{fig:2}).
\\
The phase-space trajectories in Fig.~\ref{2} are bounded by a time-dependent envelope defined by the magnitude of the complex phase-space coordinate $|z_t|$. From Eq.~(\ref{11}), it follows that the explicit field-driven term vanishes once the pulse is over, leaving the homogeneous (“free”) component, whose amplitude defines the envelope of the phase-space motion
\begin{align}
\epsilon(t) = \frac{|\mu_0' E_{0}|}{\sqrt{2\hbar\omega_0}}\Bigg|\frac{\omega_0}{2}\left[\exp(-i\varphi)J(\theta_+,t)-\exp(i\varphi)J(\theta_-,t)\right]\Bigg|.
\label{15}
\end{align}
The function $\epsilon(t)$ increases up to the pulse center $t=0$ and then decreases toward its asymptotic value. This behavior can be understood from the time derivative of $|\epsilon(t)|^2$, whose cross terms involve $J(\theta_+,t)$ and $J(\theta_-,t)$. These contributions oscillate with different frequencies and accumulate opposite phases, leading to transient destructive interference. As a result, the phase-space displacement reaches a maximum near the pulse center and subsequently decreases as the pulse vanishes.
\\
Since the vibrational dynamics is fully determined  by $z_t$, all observables can be expressed in terms of this complex variable. The instantaneous vibrational excitation is quantified by the phase-space distance $d_{\text{PS}}(t)=|z_t|$ or equivalently by $S_t=|z_t|^2$, which determines the Poisson distribution (\ref{8}) of vibrational populations.
For the driven harmonic oscillator, this also provides a direct measure of energy absorbed, since
\begin{align}
\Delta E_t = \hbar \omega_0 S_t,\label{16}
\end{align}
establishing a direct connection between classical phase-space motion and quantum excitation. We now evaluate the asymptotic excitation induced by the explicit pulse defined in Eq.~(\ref{10}). In the long-time limit, after the field has vanished, the phase-space coordinate approaches a constant value 
\begin{align}
S_\infty=\frac{\pi\sigma^2\omega_0(\mu_0'E_0)^2}{4\hbar}
\left[K_+^2+K_-^2-2K_+K_-\cos(2\varphi)\right],\label{17}
\end{align}
where $K_\pm=\exp\left[-{(\omega_0\pm\omega)^2\sigma^2}/{2}\right]$. This expression separates resonant and non-resonant contributions through the spectral overlap factors $K_\pm$, while the phase $\varphi$ controls their interference. At resonance, the counter-rotating contribution is exponentially suppressed $(K_+\ll K_-)$, and within the rotating-wave approximation,  Eq.~(\ref{17}) reduces to
\begin{align}
S_{\infty}^{\text{res,RWA}}=\frac{\pi\sigma^2\omega_0}{4\hbar}(\mu_0' E_0)^2.
\label{18}
\end{align}
The asymptotic vibrational excitation is therefore controlled by a single parameter, set by the pulse amplitude and duration. The resulting populations follow a Poisson distribution with mean $S_{\infty}^{\text{res,RWA}}$, implying that increasing the field strength directly enhances the occupation of higher vibrational levels. This behavior is illustrated in Fig.~\ref{fig:3}, which shows the corresponding increase in phase-space displacement and population spread.
Increasing $S_{\infty}^{\text{res,RWA}}$ enhances the phase-space distance, thereby increasing the energy stored in the vibrational mode, which follows from Eq.~(\ref{16}). This results in a systematic population of higher vibrational levels, as shown in  Fig.~\ref{fig:3}(b).
\\
We now analyze the vibrational populations in the non-resonant regime using Eq.~(\ref{17}). Here, the phase-space distance is strongly suppressed resulting in a correspondingly weak excitation. The dependence of the populations and phase-space distance on the detuning parameter $\theta_- = |\omega - \omega_0|$ is shown in Fig.~\ref{fig:4}.
As the detuning increases, the phase-space distance decreases and vanishes in the long-time limit. Consequently, the vibrational population remains concentrated in the ground vibrational state, reflecting inefficient energy transfer under non-resonant driving.
\\
The phase-space formalism also provides a natural point of connection with time-dependent perturbation theory (TDPT) \cite{Mukamel1995,Ratner2002, Tannor2007,Heller2018}. In the present framework, the entire dynamics is governed by $z_t$, which is defined as a time integral over the driving field [Eq.~(\ref{9})]. This shows that the relevant small parameter is not the instantaneous strength of the interaction, but its accumulated effect over the duration of the pulse. When this accumulated effect is small, the system undergoes only a weak excursion from the origin, corresponding to a small integrated interaction $\int_{-\infty}^t V_{\text{int}}(t)dt$. In this limit, the Poisson distribution reduces to its lowest-order expansion, and the present phase-space description reproduces the TDPT results for both resonant and non-resonant driving, as illustrated in Fig.~\ref{fig:5}. 
\\
As the strength of the pulse increases, the accumulated interaction grows, leading to larger phase-space distances. In this regime, higher-order contributions become important and TDPT progressively loses accuracy until it breaks down completely. By contrast, the present approach based on Gaussian wavepacket dynamics remains exact since the dynamics is exactly captured by $z_t$. At resonance, as shown in Fig.~\ref{fig:6}, this results in a substantial phase-space distance and efficient population of higher vibrational states. For non-resonant driving, however, the breakdown of TDPT occurs only at substantially larger field strengths. As shown in Figs.~S1 and S2 of the Supplementary Material \cite{SM012026}, the two approaches remain in close agreement for $E_0=0.4$ n.u. (the same field strength as in Fig.~\ref{fig:6}), whereas substantial deviations emerge upon further increasing the strength of the field to $E_0=2$ n.u. Thus, detuning suppresses the accumulated phase-space distance, requiring substantially stronger fields for TDPT to break down under non-resonant driving.
\\
In conclusion, we have presented a non-perturbative framework for vibrational excitation of a harmonic bond induced by a linear coupling to a time-dependent external field with an arbitrary pulse shape. Within this approach, the driven harmonic oscillator evolves into a displaced coherent state (a Gaussian wavepacket with a constant width) whose dynamics is fully encoded in  the complex phase-space coordinate $z_t$. This quantity directly determines the vibrational populations, which remain Poissonian at all times with mean $|z_t|^2$, reflecting the coherent-state nature of the dynamics. In contrast to the sudden approximation \cite{Burghardt2017,Burghardt2023}, the present framework describes vibrational excitation as a dynamical response to the explicit time-dependent electric field. The present formulation reduces to the results of time-dependent perturbation theory in the limit of small phase-space displacement, while remaining valid beyond this regime when perturbative expansions fail. In this sense, the phase-space distance plays the central role in quantifying vibrational excitation, and $|z_t|^2$ may be viewed as a dynamical analogue of the Huang–Rhys parameter familiar from electronic spectroscopy \cite{Rhys1950,Rabouw2015,Turner2022}, where it characterizes equilibrium displacements between electronic surfaces. The analysis further provides a transparent phase-space picture of resonant and non-resonant driving, with excitation efficiency governed by the accumulation of the field-induced phase-space displacement, establishing it as a central quantity governing vibrational excitation under arbitrary driving fields. \\
Although the present work has focused on a single vibrational mode, the formulation extends naturally to multimode harmonic systems, where the dynamics is described by a vector of phase-space displacements associated with the individual normal modes. In this case, vibrational excitation is governed by the collective evolution of the multidimensional phase-space coordinate, providing a direct route toward describing selective mode excitation and energy redistribution in polyatomic molecules. \\
The present work has employed the exact coherent-state solution of a driven harmonic oscillator, representing the harmonic limit of the more general Gaussian wavepacket framework. For anharmonic potentials, stronger fields can drive the wavepacket into regions where anharmonicity becomes important. Nevertheless, for weakly anharmonic systems with a weakly anharmonic dipole moment, the dynamics can still be described by a class of Gaussian wavepacket dynamics methods \cite{Vanicek2023}, albeit with a Gaussian of varying width. Established formalisms such as Heller's local harmonic approximation \cite{Heller1975,Heller2018} and related semiclassical Gaussian propagation schemes \cite{Karplus1990,Leok2013,Vanicek2019,Vanicek2026} therefore provide a natural extension of the present approach to weakly anharmonic systems. Within such approaches, the anharmonic potential is locally approximated by an effective quadratic polynomial about the evolving wavepacket's center, and the Hessian of this polynomial governs the evolution of the Gaussian's width. This makes the method particularly attractive for weakly anharmonic molecular vibrations, where the deviations from harmonic motion are significant enough to produce effects such as frequency shifts and intensity modulation but not so strong to cause the breakdown of the Gaussian approximation. The explicit treatment of the external electric field can then be retained, allowing the framework developed here to be extended to anharmonic vibrational excitation. Such an extension would provide a direct connection between the analytical coherent-state picture developed in the present work and more realistic molecular potentials, while retaining the computational advantages of Gaussian wavepacket dynamics. This framework for investigating anharmonic vibrational excitation, mode-selective energy redistribution, and light--matter interactions in intense and ultrafast electric fields, will be explored in future work.
\section{Acknowledgments}
The authors acknowledge stimulating discussions with Prof. David Tannor and financial support from the EPFL.
\section{Data Availability}
The data supporting the study is available within the article. 
\bibliographystyle{apsrev4-2}
\bibliography{Non-Perturbative_Harmonic_v18}
\end{document}


\title{Supplementary Material for\\ ``Non-Perturbative Vibrational Excitation by Arbitrary Electric Fields from Classical Phase-Space Dynamics''} \author{Sameernandan Upadhyayula} \author{Ji\v{r}\'i J. L. Van\'i\v{c}ek} \email{jiri.vanicek@epfl.ch} \affiliation{Laboratory of Theoretical Physical Chemistry, Institut des Sciences et Ing\'enierie Chimiques, Ecole Polytechnique F\'ed\'erale de Lausanne (EPFL), CH-1015 Lausanne, Switzerland} \date{\today} \maketitle
In this Supplementary Material, we compare the exact quantum solution (here equivalent to Gaussian wavepacket dynamics) with time-dependent perturbation theory (TDPT) for a range of electric field strengths. We focus on two quantities that directly characterize the vibrational excitation dynamics: the time-dependent phase-space distance \(d_{\mathrm{PS}}(t)\) and the asymptotic vibrational populations \(P_n(\infty)\). The comparisons below provide a systematic assessment of the agreement between the two approaches in the weak-field regime and illustrate the progressive breakdown of  TDPT as the strength of the field is increased. These results complement the discussion presented in the main text. For each field strength, the dimensionless phase-space distance \(d_{\mathrm{PS}}(t)=|z_t|\) is compared throughout the interaction with the electric field, while the final vibrational populations are compared in the long-time limit, after the field has vanished. The phase-space distance provides a direct measure of the instantaneous vibrational excitation, whereas the asymptotic populations quantify the final redistribution of population among the vibrational eigenstates. A comparison between the two methods  therefore provides a stringent test of the perturbative description.
\vspace{-0.73cm}
\setcounter{secnumdepth}{2}

\renewcommand{\thesection}{\Roman{section}}
\renewcommand{\thesubsection}{\thesection.\Alph{subsection}}

\setcounter{secnumdepth}{2}

\setcounter{section}{0}
\setcounter{subsection}{0}

\renewcommand{\thesection}{\Roman{section}}
\renewcommand{\thesubsection}{\Alph{subsection}}

\section{Comparison of Gaussian wavepacket dynamics with TDPT under non-resonant driving}
\label{sec:comparison}
\begin{figure}[H]
\centering
\includegraphics[width=1.0\linewidth, trim=0 55 0 55, clip]{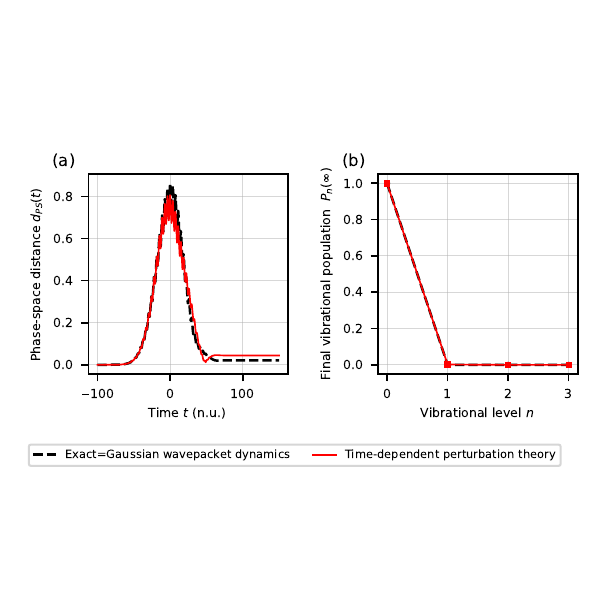}
\caption{Comparison between time-dependent perturbation theory and Gaussian wavepacket dynamics under non-resonant driving by a field of intermediate strength [same as in Fig. 6 of the main text]. (a) Time evolution of the phase-space distance. (b) Final vibrational populations.  We perform the calculations in natural units (n.u.) such that  $\mu_0'=\hbar=\omega_0=1, \ \mu_0=0$. The parameters of the electric field are as in Fig. 6 of the main text: $ \ E_0=0.4 \ \text{n.u.}, \ \omega=0.83 \ \text{n.u.}, \  \sigma=20 \ \text{n.u.}, \ \text{and} \ \varphi=0.$ } 
\label{fig:S1}
\end{figure}

\subsection{Intermediate-field non-resonant driving (Fig.~\ref{fig:S1})}
\label{sec:intermediate_nonresonant}
\begin{figure}[H]
\centering
\includegraphics[width=1.0\linewidth, trim=0 55 0 55, clip]{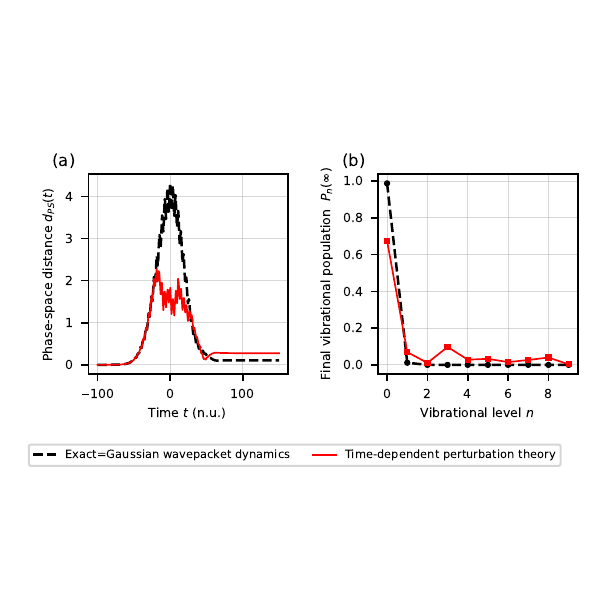}
\caption{Comparison between time-dependent perturbation theory and Gaussian wavepacket dynamics under non-resonant driving by a strong field. (a) Time evolution of the phase-space distance. (b) Final vibrational populations. The parameters used for the calculations are the same as in Fig.~\ref{fig:S1}, except $E_0=2 \ \text{n.u.}$.  } 
\label{fig:S2}
\end{figure}

We consider non-resonant driving with a detuning of
$|\omega-\omega_0|=0.17\ \text{n.u.}$ The field strength $E_0=0.4 \ \text{n.u.}$ is chosen to correspond to a regime in which TDPT breaks down completely under resonant driving, as demonstrated in Fig.~6 of the main text. Despite this field strength, the non-resonant driving produces little phase-space displacement and vibrational population transfer in both Gaussian wavepacket dynamics and TDPT (Fig.~\ref{fig:S1}). The two approaches show close agreement throughout the interaction, with only small (but discernible) deviations in the phase-space distance. Thus, although the two methods do not coincide exactly at this field strength, TDPT remains a good approximation for the non-resonant case with intermediate field strengths.
\subsection{Strong-field non-resonant driving (Fig.~\ref{fig:S2})}
To observe non-perturbative effects, we consider strong-field non-resonant driving with a field amplitude of \(E_0=2\) n.u., while retaining the same detuning as in Sec.~\ref{sec:intermediate_nonresonant}. Increasing the field strength while maintaining the same detuning does not lead to appreciable net vibrational population transfer if the quantum dynamics is performed exactly, using Gaussian wavepacket dynamics [Fig.~\ref{fig:S2}(b)]. In contrast, TDPT predicts a substantial population transfer to higher vibrational levels. A pronounced deviation is also observed in the time-dependent phase-space distance [Fig.~\ref{fig:S2}(a)], indicating that TDPT becomes unreliable in this strong-field regime, even for non-resonant driving.
\vspace{-4.7em}
\section{WEAK-FIELD $(E_0\to0)$ limit}
\begin{figure}[H]
\centering
\includegraphics[width=1.0\linewidth, trim=0 55 0 55, clip]{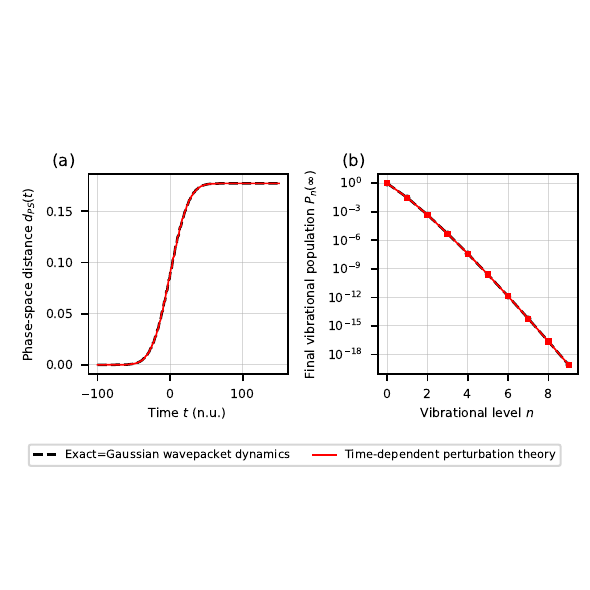}
\caption{Comparison between time-dependent perturbation theory and Gaussian wavepacket dynamics under resonant driving by a very weak field. (a) Time evolution of the phase-space distance. (b) Final vibrational populations shown on a logarithmic $y$-axis. We perform the calculations in natural units (n.u.) such that  $\mu_0'=\hbar=\omega_0=1, \ \mu_0=0$. The parameters of the electric field are  $ \ E_0=0.01 \ \text{n.u.}, \ \omega=1 \ \text{n.u.}, \  \sigma=20 \ \text{n.u.}, \ \text{and} \  \varphi=0$. } 
\label{fig:S3}
\end{figure}

\begin{figure}[H]
\centering
\includegraphics[width=1.0\linewidth, trim=0 55 0 55, clip]{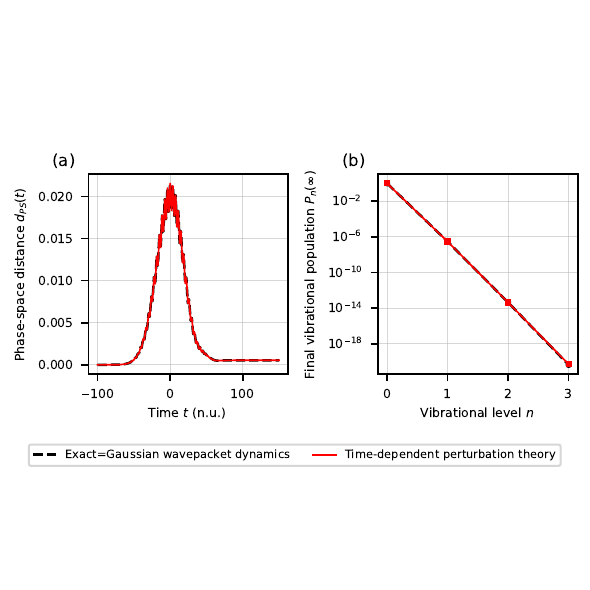}
\caption{Comparison between time-dependent perturbation theory and Gaussian wavepacket dynamics under non-resonant driving by a very weak field. (a) Time evolution of the phase-space distance. (b) Final vibrational populations shown on a logarithmic $y$-axis. The parameters used for the calculations are the same as in Fig. S3, except $\omega=0.83 \ \text{n.u.}$  } 
\label{fig:S4}
\end{figure}

Figures~\ref{fig:S3} and \ref{fig:S4} examine the $E_0\to 0$ regime for both resonant and non-resonant driving, respectively, using $E_0=0.01$ n.u. In both cases, the results obtained from TDPT and Gaussian wavepacket dynamics are nearly identical. The phase-space distance \(d_{\mathrm{PS}}(t)\) and the asymptotic vibrational populations \(P_n(\infty)\) show perfect agreement for both resonant and non-resonant driving. These results confirm that TDPT correctly reproduces the exact Gaussian wavepacket dynamics in the $E_0\to0$ limit for both resonant and non-resonant driving.
